\documentclass[12pt,a4paper]{article}
\usepackage{amsmath}
\usepackage{bm}
\usepackage{amsfonts}
\usepackage{graphicx}
\usepackage[normal]{caption2}
\usepackage{subfigure}
\usepackage{rotating}
\usepackage{citesort}
\usepackage{lscape}
\usepackage{section}
\begin{document}
\renewcommand\arraystretch{1.1}
\setlength{\abovecaptionskip}{0.1cm}
\setlength{\belowcaptionskip}{0.5cm}
\title {The study of fusion barriers of neutron-rich colliding nuclei using various isospin-dependent potentials}
\author {Maninder kaur \\
 \it
\it Applied Sciences Department, \\\it RBIEBT, \it Mohali Campus, INDIA\\
} \maketitle
PACS number: 24.10.-i Nuclear reaction models and
methods 25.70.Jj Fusion and fusion-¯ssion reactions 25.60.Pj-
Fusion reactions { 25.70.-z intermediate energy heavy-ion reactions \\
 Electronic address:~maninderphysics@gmail.com

\begin{abstract}
A detailed study fusion of neutron-rich colliding nuclei is
performed using various isospin dependent potentials. For present
study, Three different series namely, Ne-Ne, Ca-Ca, and Zr-Zr are
taken into account and N/Z ratio.A monotonous increase (decrease)
in the fusion barrier positions (heights) using a unified second
order nonlinear parametrization in the normalized fusion barrier
positions and heights with (N/Z-1) is presented. These
predications are in good agreement with the available theoretical
as well as experimental results.
\end{abstract}
\section{Introduction}
 \par
 With the availability of radioactive-ion nuclear beams the fusion
of colliding nuclei with the excess of neutron/proton ratio or
near the drip line has attracted central position in the current
days research \cite{1,2,3}.The neutron -rich radioactive-ion beams
have been applied to synthesize new, neutron -rich heavy nuclei.
This is because in the synthesis of heavy nuclei with neutron-rich
projectile, one expect a higher survival probability of the
completely fused system due to its lower fissility and lower
excitation energies. once we go beyond binding energies of
colliding nuclei many more phenomena like collective flow, multi
fragmentation, stopping and sub-barrier particle production also
appeared as dominant modes \cite{4,5,6,7,8}.
\par
Several experiential studies have also been undertaken in the
literature to study the effect of varying the N/Z -ratio of the
projectile and target nuclei upon the fusion cross sections
\cite{6,7,10,11}. In addition, the fusion transfer at the neck
region has been suggested \cite{12,13}. A microscopic description
of the formation of neck in the fusion reaction remain a challenge
to microscopic theories. The influence on sub-barrier fusion of
processes such as, transfer \cite{14} and breakup reaction
\cite{9,15} is not yet clear; moreover, the effect of unusual
structure, such as halos and skins \cite{16}, is being studied
\cite{7,17}. Recently, Sun et al. \cite{18}, suggested that N/Z
may be used as an experimental observable to extract neutron skin
for neutron rich nuclei. All above experimental as well as
theoretical information indicates that the dynamics of neutron/
proton -rich nuclei is not fully understood and needs further
attention.
\par
The properties of various neutron -rich nuclei with different N/Z
-ratio are studied in the literature
e.g.:$^{9-10}$$_{2}$He(N/Z=3.50-4.00; where N and Z are the
neutron and proton content of the nucleus),
$^{6,8,9,11}$$_{3}$Li(N/Z=1.0,1.67,2.0,2.67),
$^{22}$$_{6}$C(N/Z=2.67), $^{26-28}$$_{8}$O(N/Z=2.25-2.50),
$^{31}$$_{9}$F(N/Z=2.444),
$^{28,32,34}$$_{10}$Ne(N/Z=1.8,2.2,2.40),
$^{30-32,37}$$_{11}$Na(N/Z=1.727-1.909,2.364),
$^{40}$$_{12}$Mg(N/Z=2.333),
$^{49-51}$$_{18}$Ar(N/Z=1.722-1.8333), $^{60}$$_{20}$Ca(N/Z=2.0),
$^{57-60}$$_{25}$Mn(N/Z=1.28-1.4),
$^{68-78}$$_{28}$Ni(N/Z=1.429-1.786),
$^{84,86}$$_{30}$Zn(N/Z=1.8,1.87),
$^{90,92}$$_{32}$Ge(N/Z=1.813,1.875), $^{132}$$_{50}$Sn(N/Z=1.64),
$^{123}$$_{47}$Ag(N/Z=1.617),
$^{123-128}$$_{48}$Cd(N/Z=1.563-1.667)[4,32] and proton-rich are
$^{6}$$_{4}$Be(N/Z=0.50), $^{10}$$_{7}$N(N/Z=0.429),
$^{12}$$_{8}$O(N/Z=0.50), $^{17}$$_{9}$F(N/Z=0.89),
$^{22}$$_{14}$Si(N/Z=0.571), $^{31}$$_{18}$Ar(N/Z=0.722),
$^{34}$$_{20}$Ca(N/Z=0.70), $^{38,39}$$_{22}$Ti(N/Z=0.727,0.773),
$^{45,49}$$_{26}$Fe(N/Z=0.731,0.885),
$^{48,49,53}$$_{28}$Ni(N/Z=0.714-0.75,0.893),
$^{54}$$_{30}$Zn(N/Z=0.80),
$^{217}$$_{92}$U(N/Z=1.359)etc.\cite{19,20,21}.
\par
A suitable set of models are therefore needed to study the
dynamics of neutron/proton -rich colliding nuclei. A large number
of theoretical models are available in the literature based upon
the different assumptions \cite{1,2,3}. Among them, proximity
potential due to Blocki et al. \cite{22}, is well known for its
simplicity and wider applications in different fields. Several
modifications or refinements over the original proximity potential
are also available in the recent time by including either
up-to-date knowledge of the surface energy coefficient or nuclear
radii \cite{3,23}. Various authors, modified or parameterized
their approaches within the proximity concept \cite{3}. All these
modifications include new emerging degree of freedom i.e. isospin,
either in radius formula, universal function and/or in surface
energy coefficient \cite{3}. Further, The outcome will definitely
different if one use such type of models in the isospin plane.
\par
Recently, one of us and collaborator, have carried out a detailed
study involving symmetric as well as asymmetric colliding nuclei
using 16 proximity-type potentials \cite{3}. Unfortunately, the
maximum N/Z content of all the experimentally studied heavy-ion
reactions is 1.60 (i.e.$^{6}$He+$^{238}$U) \cite{7}. On the other
hand, the first measurement with the proton drip line nucleus is
of $^{17}$F + $^{282}$Pb (with N/Z=1.473) \cite{20}.Therefore, a
systematic study of nuclei having larger N/Z ratio using new
proximity-type potentials is in demand. Further, it gives us a
unique possibility to test the validity or accuracy of these
models for the nuclei far away from the line of stability.
Therefore, a systematic dependence of fusion barrier (heights and
positions) and cross sections using various isospin dependent
models on neutron excess is needed. Similar study was also
presented by Puri et al. \cite{1}, where only Ca and Ni series
were used. In the present study, we extend the work to include new
series like Ne-Ne (with N/Z ratio = 0.6-2.0) and Zr-Zr (with N/Z =
0.75-2.0) along with Ca-Ca (with N/Z = 0.5-2.0) series. The
overall domain of N/Z ratio is from 0.5 to 2.0 for all series. The
asymmetry parameter A$_{s}$ (N/Z-1) of the colliding nuclei varies
between -0.5 and 1.0. Note that non zero value of A$_{s}$ will
involve complex interplay of the isospin degree of freedom which
has strong role at intermediate energies as well \cite{4}. Section
2, deals with fine points of the models in brief, Section 3
contains the results and summary is presented in Section 4.
\par
\section{The Model}
\par
The total ion-ion interaction potential V$_{T}$(r) comprises of
nuclear and Coulomb part:
\begin{equation}
V_{T}(r) = V_{N}(r)+V_{C}(r).
\end{equation}
Here V$_{C}$(r) = Z$_{1}$Z$_{2}$e$^{2}$/r is a good approximation,
because fusion happens at a distance greater than touching
configuration of the colliding pair.
\par
The nuclear part of the ion-ion interaction potential V$_{N}$(r)
is calculated within the proximity concept. All proximity
potentials are based upon the proximity force theorem, according
to which \cite{22}, "\emph{the force between two gently curved
surfaces in close proximity is proportional to the interaction
potential per unit area between the two flat surfaces}". In
original proximity potential \cite{22}, the nuclear part of the
interaction potential V$_{N}$(r) can be written as
\begin{equation}
V_{N}(r) = 4\pi\gamma b\bar{C}\phi(\frac{r-C_{1}-C_{2}}{b})\textrm
 { MeV}.
\end{equation}
In this, $\overline{C}$ is the reduced radius with equivalent
sharp radius R$_{i}$ as
\begin{equation}
R_{i}=1.28A^{1/3}_{i}-0.76+0.8A^{-1/3}_{i} \textrm{fm} (i=1,2)
\end{equation}
where $\phi$ $(\xi=\frac{r-C_{1}-C_{2}}{b})$is the universal
function that depends on the separation between the surfaces of
two colliding nuclei only. Both these factor do not depend on the
isospin content. However, the last parameter $\gamma$, the surface
energy coefficient, depends upon the neutron/proton excess as

\begin{equation}
\gamma=\gamma_{0}[1-K_{s}(\frac{N-Z}{A})^{2}]
\end{equation}


Where N, Z being the total number of neutrons and protons. In the
original version, $\gamma$$_{0}$ = 0.9517 MeV/fm$^{2}$ and k$_{s}$
= 1.7826 \cite{22}. Noted that for symmetric nuclear matter, N =
Z, $\gamma$ = $\gamma$$_{0}$ = 0.9517 MeV/fm$^{2}$ indicating
maximum strength of the potential. If we move to neutron (proton)
-rich colliding nuclei with N$>$ Z (N $<$ Z) then $\gamma$ starts
decreasing resulting in comparatively lesser attractive potential.
Later on, these coefficients were further improved by M\"{o}ller
and Nix with values $\gamma$$_{0}$ = 1.2496 MeV/fm$^{2}$ and
k$_{s}$ = 2.3 \cite{24}. This is labelled as Prox 88 \cite{2,25}.
In the latest version of proximity potential \cite{23}, $\gamma$
has form based on the precise neutron skin as
\begin{equation}
\gamma=\frac{1}{4\pi r_{0}^{2}}[18.63(\textrm{MeV}
)-Q\frac{(t_{1}^{2}+(t_{2}^{2})}{2r_{0}^{2}}].
\end{equation}

\par
The corresponding proximity potential is labelled as Prox 00
\cite{3}. One of us and collaborator \cite{3}, modified above
potential to include latest radius formula \cite{26} and is
denoted as Prox 00N. Note that both Prox 00 and Prox 00N has
isospin dependent radius with slightly different constants whereas
the factors surface energy coefficient $\gamma$ and the universal
function $\phi$(s) are same. In both newer versions of Bass
(labelled as Bass 77 and Bass 80 in Ref. \cite{3}) radius is
slightly changed to
\begin{equation}
R_{i}=1.16A^{1/3}_{i}-1.39A^{-1/3}_{i} \textrm{fm} (i=1,2)
\end{equation}



and then to sharp radius as is used in Prox 77 in the later
version (i.e. Bass 80). Both newer versions of Winther (marked as
BW 91 and AW 95 in Ref. \cite{3}) has again similar expression for
$\gamma$ as is given in Prox 77 with slight difference that here
isospin content is calculated separately for the target/
projectile. Whereas, first version due to Winther (labeled as CW
76 in Ref. \cite{3}) does not have any $\gamma$ dependence. Even
radii are function of mass only. Both versions of Ng\^{o} (labeled
as Ng\^{o} 75 and Ng\^{o} 80 in Ref. \cite{3}) do not consider
$\gamma$, but latest version of Ng\^{o}  (Ng\^{o} 80) has isospin
dependence in radius parameter. On the other hand, a complex
isospin dependence in the universal function $\phi$(s) and radius
is given in the version of Denisov \cite{27}. Also by using the
latest form of radius given in Ref. \cite{26} in Denisov potential
resulting in closer agreement with the experimental data for
fusion barrier heights and cross-sections. This potential is
labeled as Denisov N \cite{3}. All the above mentioned
proximity-type potentials are able to reproduce the experimental
fusion barriers within $\pm$10 on the average \cite{3}.
\par
In total, 8 proximity-type potentials are used in the present
study. Among them, three are basic proximity potentials (Prox 77,
Prox 88, and Prox 00), three due to Bass ( Bass 80), Winther (AW
95) and Ng\^{o} (Ng\^{o} 80) each, and two newly modified
potentials (Prox 00N and Denisov N) are used. The model due to
Bass et al., (Bass 80) used in the present analysis is independent
of isospin dependence. For the detail of the models reader is
refer to Ref. \cite{3}. In most of the potentials and versions,
modifications are made either through the surface energy
coefficients or in nuclear radii. These two rather being technical
parameters can have sizeable effects on the outcome of a reaction
\cite{3}.
\par
 From these brief outlines, it is clear that much stress is
made on the surface energy coefficients $\gamma$ and nuclear radii
to incorporate the isospin factor of a potential. Definitely,
based on different assumptions and isospin dependence, different
versions will respond to the collision of neutron -rich or
-deficient nuclei differently compared to N = Z nuclei.
\par

\section{Results and Discussion}
The present study deals with large variety of above mentioned
potentials. Using these potentials, we firstly calculate the total
ion-ion interaction potential using Eq. (1). Once total ion-ion
interaction potential is calculated, one can extract the barrier
height V$_{B}$ and barrier position R$_{B}$ using conditions:
\begin{equation}
\frac{dV_{T}(r)}{dr}\mid _{r=R_{B}} =0;
and\frac{d^{2}V_{T}(r)}{dr^{2}}\mid _{r=R_{B}}\leq0.
\end{equation}

\par
Here we consider the collisions of three different series namely;
$^{A1}Ne$+$^{A2}Ne$(with N/Z = 0.6 to 2.0);
$^{A1}Ca$+$^{A2}Ca$(with N/Z = 0.5 to 2.0), and
$^{A1}Zr$+$^{A2}Zr$ (with N/Z = 0.75 to 2.0) to cover wider mass
range. We starts with the collision of N = Z nuclei and then add
(or remove) neutrons gradually from either of the colliding pairs
till we reach N = 2Z (N = 0.5Z) nuclei. In total, 150 such
collisions involving different isotopes of different series are
taken into account.
\par
As a first step, we check the effect of addition or removal of
neutrons on the nuclear part of the interaction potential
V$_{N}$(r) in different models. In Fig. 1, we display V$_{N}$(r)
as a function of internuclear distance r for the reactions of
$^{16}Ne$+$^{16}Ne$, $^{16}Ne$+$^{20}Ne$, $^{20}Ne$+$^{20}Ne$,
$^{20}Ne$+$^{28}Ne$ and $^{28}Ne$+$^{28}Ne$ using eight proximity
type potentials. Based upon the different assumptions used in
different models, shape as well as strength of the potential
differ accordingly. From this plot we can compare the different
models to check the isotopic dependence of the interaction
potential. from this plot it is evident that the V$_{N}$ becomes
deeper with the addition of neutrons whereas the reverse is true
for the removal of neutrons. At the same time, the general shape
at the surface region is same. In particular, Bass 80, Prox 00,
and Prox 00N have no repulsive core at shorter distances, whereas
AW 95 follow Woods-Saxon type form. On the other hand, Ng\^{o} 80,
Denisov N, Prox 77, and Prox 88 have repulsive core at shorter
distances. In addition, four effects of addition (removal)
neutrons to N = Z nuclei are clearly visible: (i) barrier height
is decreased (ii) barrier position is increased (iii) depth of the
potential is increased in all potentials except Denisov N, and
(iv) diffuseness of the potential is also changes. These effects
will definitely influence the fusion probability at below barrier
energies. Therefore, before discussing the enhancement of the
fusion cross section for the neutron -rich fusion reaction, we
investigate the systematic dependence of fusion barriers on the
neutron asymmetry parameter A$s$(= N/Z-1). Using the above sets of
models, fusion barrier heights and positions are calculated for
150 colliding pairs using 8 sets of models.
\par
Here the proton-rich systems show the deeper pocket compared to
neutron-rich systems. This is due to the reason that the form of
radius used in Denisov N has very complex dependence on the mass
number A. In Fig. 2. we have plotted V$_{N}$ as a function of
internuclear distance r, but for Zr-Zr reactions. In this case
again Same results are obtained as in the case of Ne-Ne. Again
Denisov N shows exceptional behavior. These two graphs basically
shows the same trend with the addition of neutrons/protons. As we
move from the lighter systems to heavier ones i.e., from Ne to Zr
series, the scattering around the mean values decreases. this
implies that as we move from the lighter to heavier systems, the
role of neutron content diminishes.
\par
 The total potential containing the nuclear and coulomb parts will
 also be affected by the neutron contents.In Fig. 3. we have
 plotted total potential V$_{T}$(r) as a function of internuclear
 distance r for Zr-Zr colliding pairs . With the addition
 of neutron barrier height decreases and the barrier position increases.
 Whereas the reverse is true for the removal of neutrons. As a
 result, the fusion probability increases with the addition of
 neutrons. Again some discrepancies have been noticed for Denisov
 N in the repulsive part of potential. Similarly, we have plotted
 the total potential V$_{T}$ as a function of r for the Ne-Ne and
 Ca-Ca (not shown here). from these plots we observed that as we
move from the lighter to heavier ones i.e., from Ne to Zr series,
explicit mass dependence is more visible. This indicates that the
conclusion based on island of the periodic table can be
 misleading.
 \par
 For a model independent analysis, we see the same effect in all
 these models. Some differences in various potentials have been
 noticed at the surface regions. We have noticed that the modal
 ingredients such as nuclear radii, reduced radius, surface energy
 coefficient and universal function, have sizable effect on the
 interaction potential as well as on the fusion probability. We
 can simply say that different potential use different radii,
 surface energy coefficient and universal function leading to
 different mass dependence. Some of these reactions along with
 barrier heights and barrier positions are summarized in tables.
 In these Figs., we have observed that Bass 80, Prox 00, and
 Prox 00N do not have any pocket. This is due the reason that Prox
 00 and Prox 00N are derived only for the distances greater that
 touching configuration and Bass 80 is based on the classical
 assumptions. All other potentials have pockets. We have also
 notices that the neutron/proton content also affect the
 diffuseness of the pocket. The nuclear interaction potential is
 more attractive for Prox 88 compared to other potentials. The
 shape of the different potentials is different because according
 to Proximity theorem, The nuclear part of the potential i.e.,
 V$_{N}$(r) depends on $\phi$(s) and $\overline{C}$. And the shape
 of the potential depends on the universal function $\phi$(s). All
 the potentials, except Bass 80, have isospin dependence in
 $\overline{C}$. As a result, the contribution of $\overline{C}$
 is different in all potential. Also we have noticed that the
 contribution of $\overline{C}$ due to isotopic dependence is
 stronger for the proton-rich colliding nuclei compared to
 neutron-rich nuclei. Also the change in the neutron content also
 affect the depth of the pocket of the potential. We have observed
 that the pocket is less deep in the case of proton-rich nuclei.
 This implies that these reactions are less favorable for the
 fusion of colliding nuclei.
 \par
 In Fig. 4., We have used only three versions of
 Proximity potential these versions differ due to different forms
 of $\gamma$ . Here the contribution due to the surface energy
 coefficient, $\gamma$ at barrier heights and positions is taken.
 Alternatively, we can say that the isotopic dependence of any
 quantity can examined more clearly by plotting different quantity
 against asymmetric term A$_{S}$. We have observed that if we
 start from N=Z i.e., symmetric system and add or remove neutrons
 from symmetric systems the contribution of $\gamma$ decreases on
 the both sides of symmetric point. The contribution of $\gamma$
 is much stronger in the case of proton rich nuclei as compared to
 the neutron rich nuclii.the contribution of $\gamma$ towards the
 Prox 88 potential is much stronger compared to remaining two
 versions. Similarly in Fig. 5.,we  display the universal function
 $\phi$ (s), calculated at barrier position, versus A$s$ for all
 the potential. The variation of $\phi$ (s) is smooth throughout
 the variation of neutron content in the case of AW 95 and Denisov
 N. Whereas Bass 80 and Ng\^{o} 80 show slight isotopic dependence
 as we move away from the N=Z symmetric  line. The variation in $\phi$
 (s) is different for neutron neutron-rich nuclei compared to
 proton-rich nuclei. This variation of $\phi$ (s) is more for
 neutron-deficient nuclei whereas the variation of $\phi$ (s) is
 almost saturates in the case of neutron-rich colliding nuclei. As
 we move from lighter to heavier system e.g., from Ne-Ne to Zr-Zr,
 all the potential converge to same results as shown in Fig. 5.
 indicating the mass independent observation.
 \par
The variation in the fusion barrier positions with neutron/proton
content is analyzed in Figs. 6 and 7. Here, we display the
variation of $\triangle$ R$_{B}$(\%)  and $\triangle$ V$_{B} $(\%)
defined as
\begin{equation}
\triangle R_{B}(\%)=\frac{R_{B}-R_{B}^{0}}{R_{B}^{0}}\times100,
\end{equation}
\begin{equation}
\triangle V_{B}(\%)=\frac{V_{B}-V_{B}^{0}}{V_{B}^{0}}\times100,
\end{equation}
\par
as a function of asymmetry parameter A$_{s }$(= N/Z-1) using eight
sets of potentials discussed above. Here, R$_{B}$$^{0}$ and
V$_{B}$$^{0}$ are, respectively, the fusion barrier position and
height corresponding to (N = Z) colliding nuclei and R$_{B}$ and
V$_{B}$ refer for neutron/proton -rich colliding nuclei. The main
advantage of these normalized variation is that it gives mass
independent picture. For the present picture, we starts with the
collision of N=Z nuclei, then gradually add/remove neutrons from
either of the colliding pair. For example, we started with the
collision of $^{40}Ca$+$^{40}Ca$ , then add neutrons gradually, by
keeping charges Z1 and Z2 always fixed. In this series, at the end
of the chain we have the collision of $^{60}Ca$+$^{60}Ca$.
Similarly, if we remove the neutrons then we have at the end the
collision of $^{30}Ca$+$^{30}Ca$. It is clear from the figures,
that all the models follow a unified non-linear second order
parametrization given as:
\begin{equation}
\triangle R_{B}(\%)=a(\frac{N}{Z}-1)+b(\frac{N}{Z}-1)^{2},
\end{equation}
\begin{equation}
\triangle V_{B}(\%)=c(\frac{N}{Z}-1)+d(\frac{N}{Z}-1)^{2},
\end{equation}
Here, a, b, c, and d are the constants varies from model to model
and its values are displayed in Figs 2 and 3. The above results
are in agreement with the recent work due to Puri et al., for Ca
and Ni series \cite{1}, The available experimental as well as
theoretical data is also displayed. All the available theoretical
as well as experimental data follow our parametrization pattern
very well except few points due to experimental uncertainty in
different experimental setups. The theoretical data is taken from
Refs. \cite{28}, whereas, the experimental data is taken from
Refs. \cite{29}, The above pattern indicates that, with the
addition of neutron to N = Z nuclei, fusion barrier position is
increased and barrier height decreased, whereas, reverse happen
for neutron -deficient nuclei. All models do not show much
scattering from the middle curve as one move from N = Z to very
neutron -rich and -deficient colliding nuclei. Bass 80 and Ng\^{o}
80 show slight scattering for neutron -rich and -deficient nuclei.
It may be due to the reason that the isospin dependence included
in Ng\^{o} 80 model have not much effect on N/Z ratio, whereas,
Bass 80 is independent of such kind of dependence. We further note
that the slopes of the central line also varies from model to
model, whereas, the overall pattern is nicely explained by the
above parametrization. It may be due to the different assumptions
used in different models.
\par
Along with $\triangle$ R$_{B}$(\%) and $\triangle$ V$_{B}$(\%), we
also studied $\triangle$ V$_{N}$(\%) and $\triangle$V$_{C}$(\%)
using the same set of models and series (not shown here). We see
that these variations show larger scattering due to the different
assumptions resulting from different forms of radii, universal
function, surface energy coefficients etc. The diffuseness
parameter `a' is also different in some potentials.
\par
In Fig. 8. and Fig. 9., we have plotted $\triangle$ V$_{C}$(\%)and
$\triangle$ V$_{N}$(\%) as a function of asymmetry parameter A$_{s
}$(= N/Z-1). These normalized quantities are given by:
\begin{equation}
\triangle V_{C}(\%)=\frac{V_{C}-V_{C}^{0}}{V_{C}^{0}}\times100,
\end{equation}
\begin{equation}
\triangle V_{N}(\%)=\frac{V_{N}-V_{N}^{0}}{V_{N}^{0}}\times100,
\end{equation}
\par
 In the case of neutron-deficient nuclei, not only the nuclear
 potential becomes more attractive, but at the same time, the
 Coulomb forces become stronger, therefore, their mutual dominance
 decides about the fate of the barrier. It is clear from the Fig.
 8. that the increase in Coulomb potential is much more compared
 to the corresponding nuclear potential, therefore, enhancing the
 fusion barriers when neutron are removed. Nuclear potential is
 different for different colliding series. Nuclear potential that
 also includes geometrical factor has a monotonic isotopic removal
 of neutrons. This result is in contradiction to the couple of
 results calculated earlier where it was discussed that the
 nuclear part of the potential is more attractive with addition of
 neutrons and leading to the reduced barrier. In general, we can
 say that all the different models converge to nearly same
 results. More experiments are needed to verify our prediction. A
 considerable mass dependence is seen in the case of Bass 80 and Ng\^{o}
 80. Aw95 and Denisov N also show slight mass dependence but Prox
 77, Prox 88, Prox 00 and Prox 00N potentials indicate a mass
 independent isotopic effect in fusion dynamics. Similarly in Fig.
 9. large scattering is observed in all the cases. Again here Bass
 80, Ng\^{o} 80, AW95 and Denisov N show large scattering compared
 to Prox 77, Prox 88, Prox00 and Prox 00N potentials.

\begin{figure}[!t]
\centering
 \vskip 1cm
\includegraphics[angle=0,width=12cm]{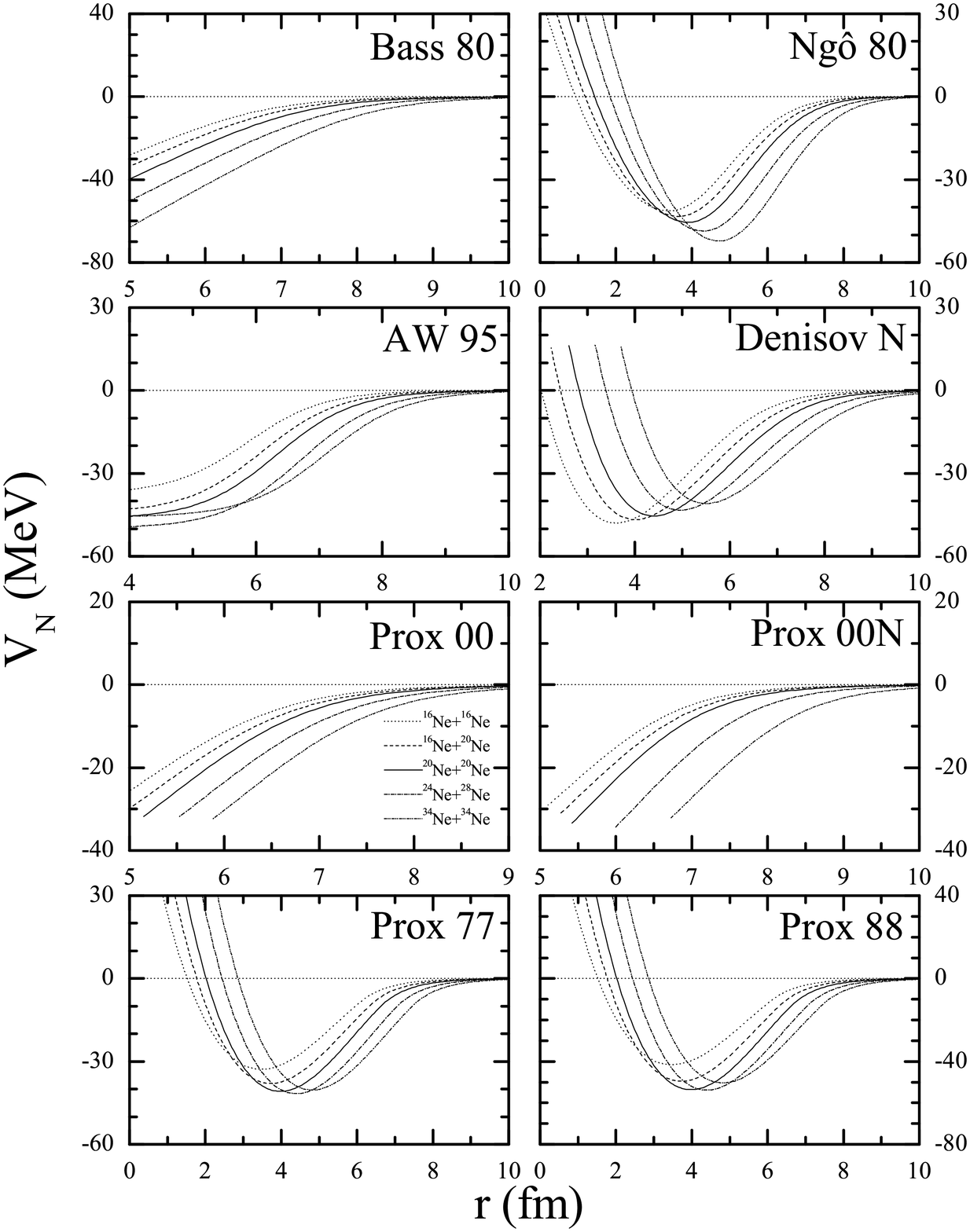}
 \vskip -0cm \caption{The nuclear part V$_{N}$(MeV) as a function of
internuclear distance r for the reactions of
$^{16}Ne$+$^{16}Ne$,$^{16}Ne$+$^{20}Ne$,$^{20}Ne$+$^{20}Ne$,$^{20}Ne$+$^{28}Ne$
and $^{28}Ne$+$^{28}Ne$  using eight sets of proximity-type
potentials.}\label{fig1}
\end{figure}

\begin{figure}[!t]
\centering
 \vskip 1cm
\includegraphics[angle=0,width=12cm]{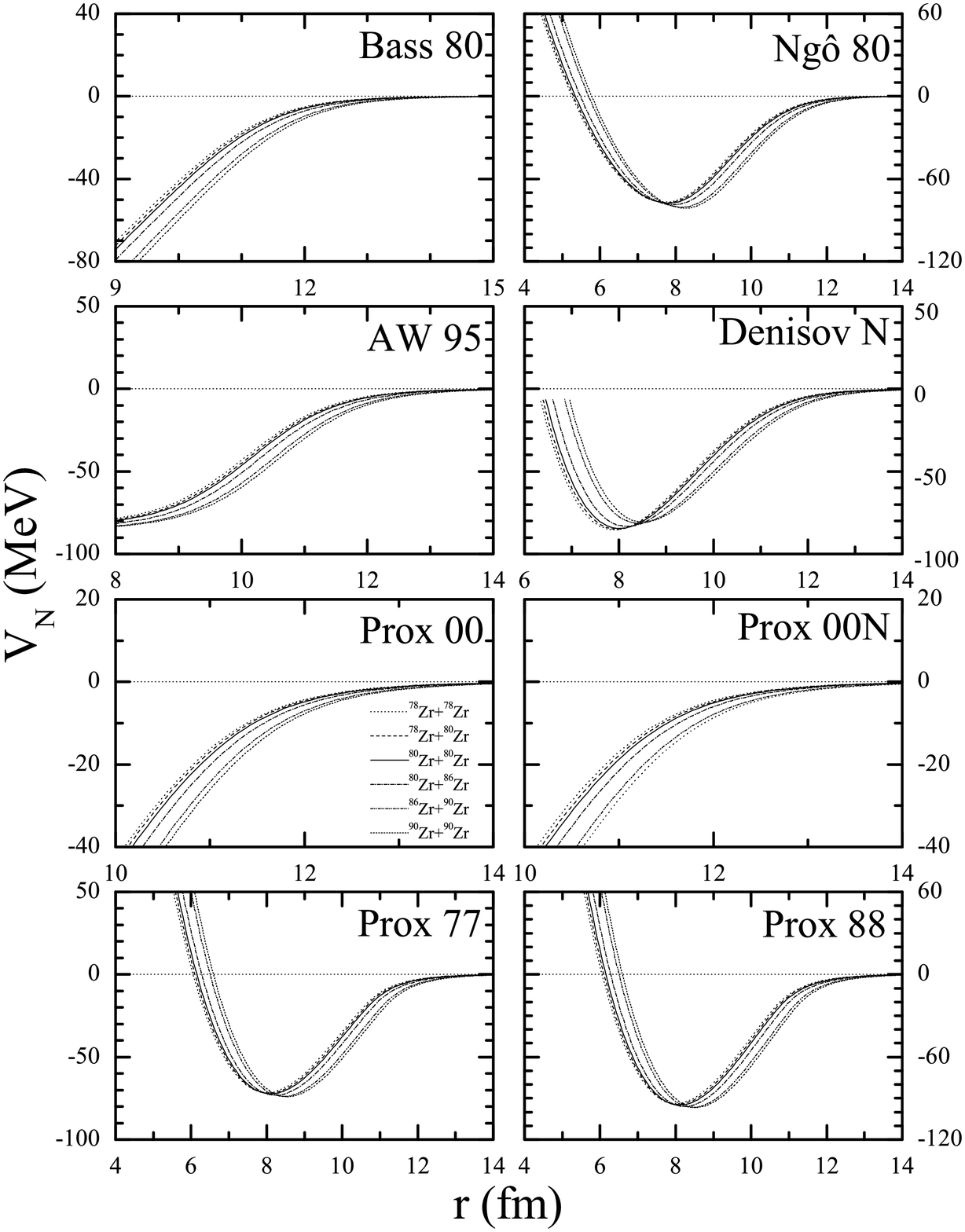}
 \vskip -0cm \caption{ Same as fig1, but it is plotted for Zr-Zr}\label{fig1}
\end{figure}

\begin{figure}[!t]
\centering
 \vskip 1cm
\includegraphics[angle=0,width=12cm]{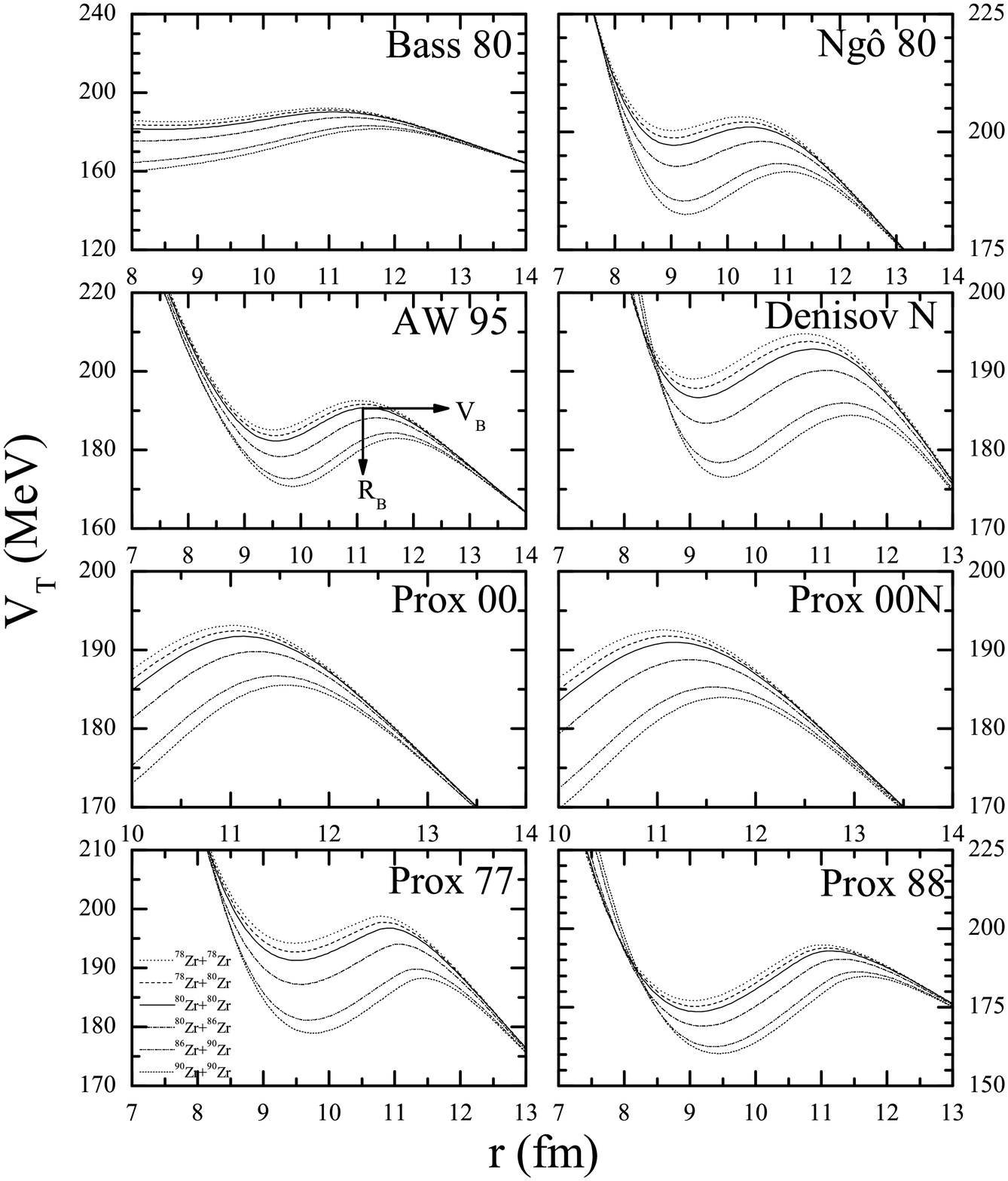}
 \vskip -0cm \caption{The total interaction potential V$_{T}$(r)(MeV) is
plotted as a function of internuclear distance r for Ne-Ne
colliding pairs using eight different potentials. Here neutron as
well as proton-rich colliding pairs are considered. For one pair
we also display barrier height V$_{B}$ and position
R$_{B}$}\label{fig1}
\end{figure}

\begin{figure}[!t]
\centering
 \vskip 1cm
\includegraphics[angle=0,width=12cm]{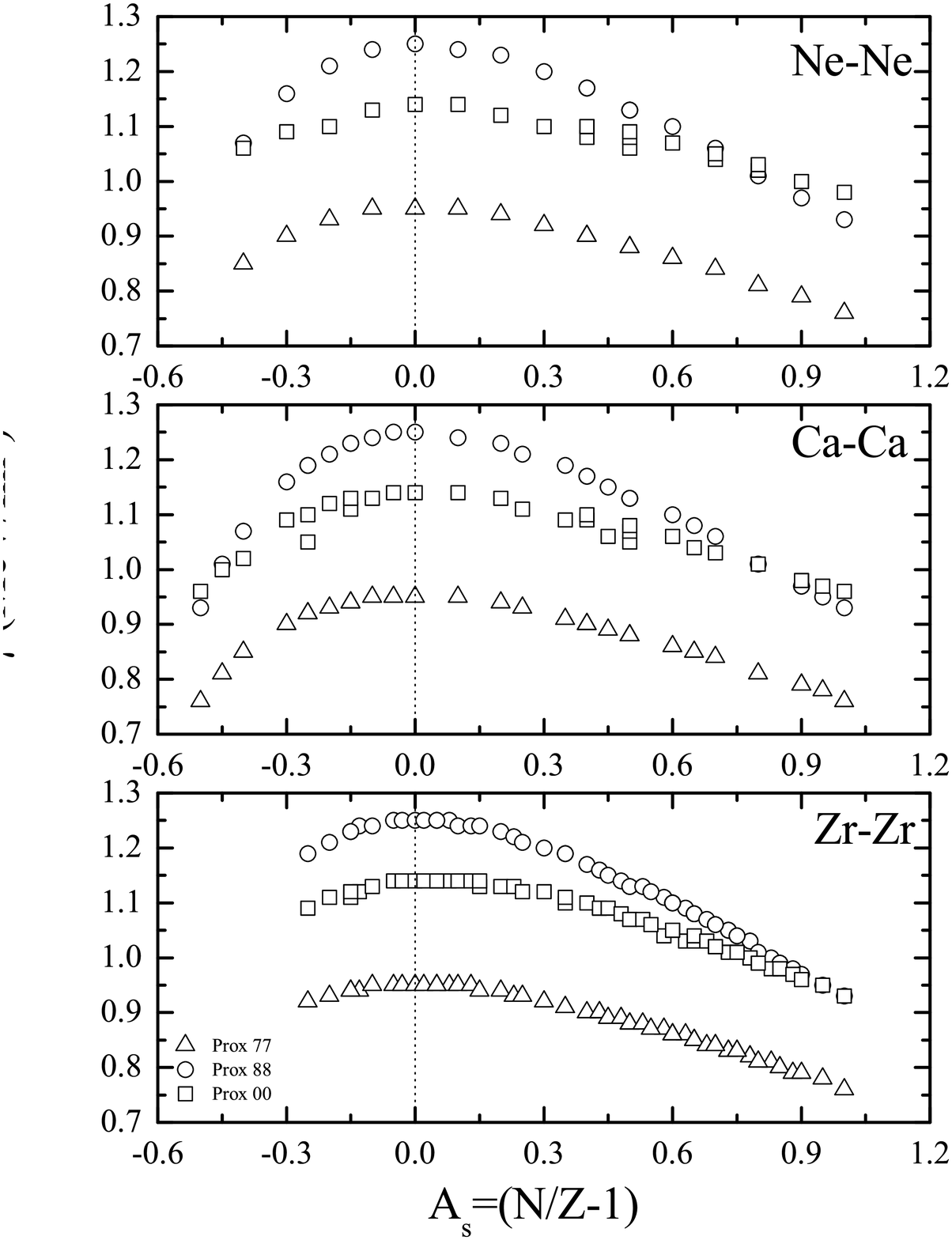}
 \vskip -0cm \caption{Variation of surface energy coefficient
$\gamma$(MeV/fm$^{2}$) as a function of asymmetry parameter for
three different series using three versions of Proximity
potentials.  }\label{fig1}
\end{figure}

\begin{figure}[!t]
\centering
 \vskip 1cm
\includegraphics[angle=0,width=12cm]{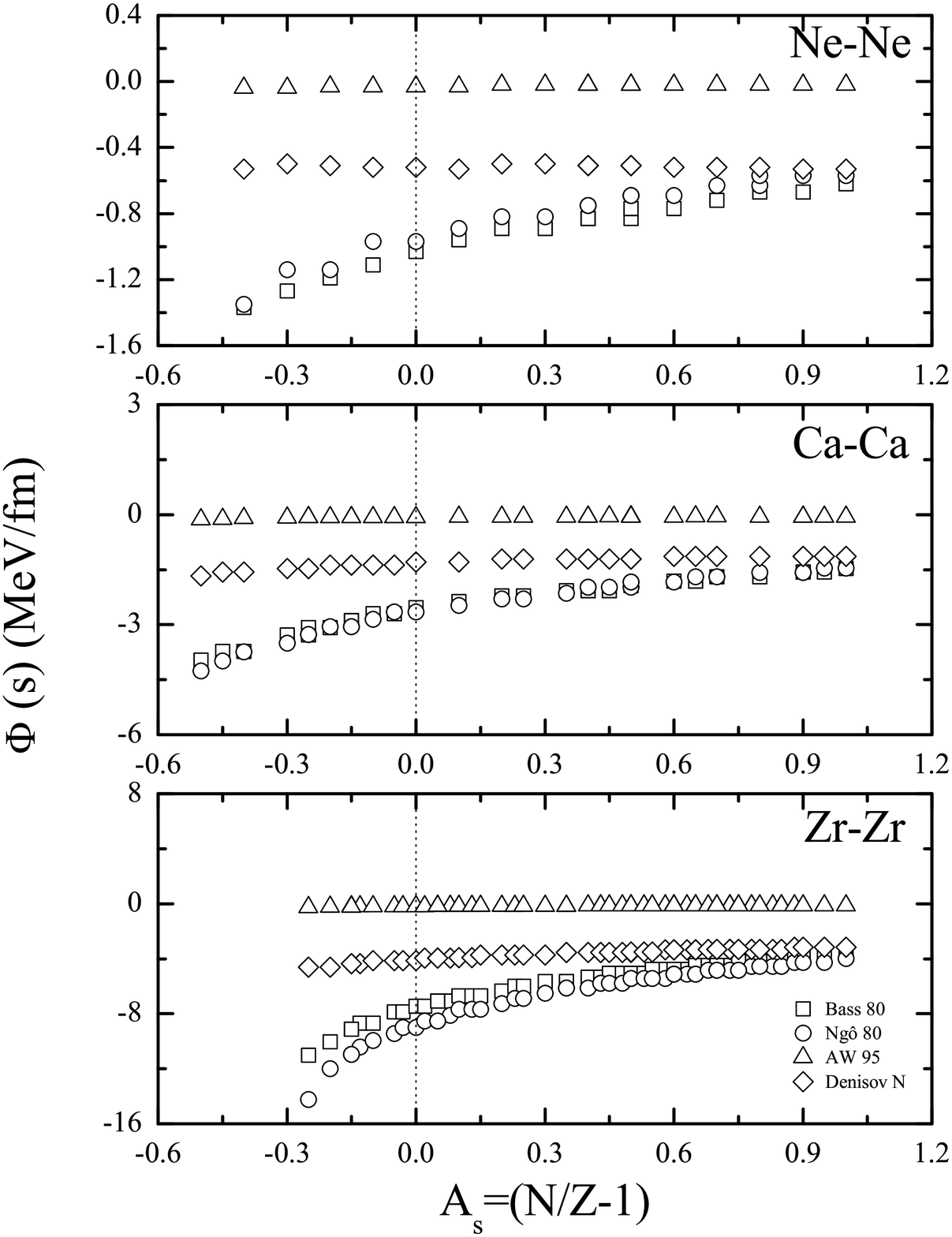}
 \vskip -0cm \caption{Variation of universal function $\phi$ as a function of
asymmetry parameter A$_{S}$ for three different series using four
versions of Proximity potentials.}\label{fig1}
\end{figure}

\begin{figure}[!t]
\centering
 \vskip 1cm
\includegraphics[angle=0,width=12cm]{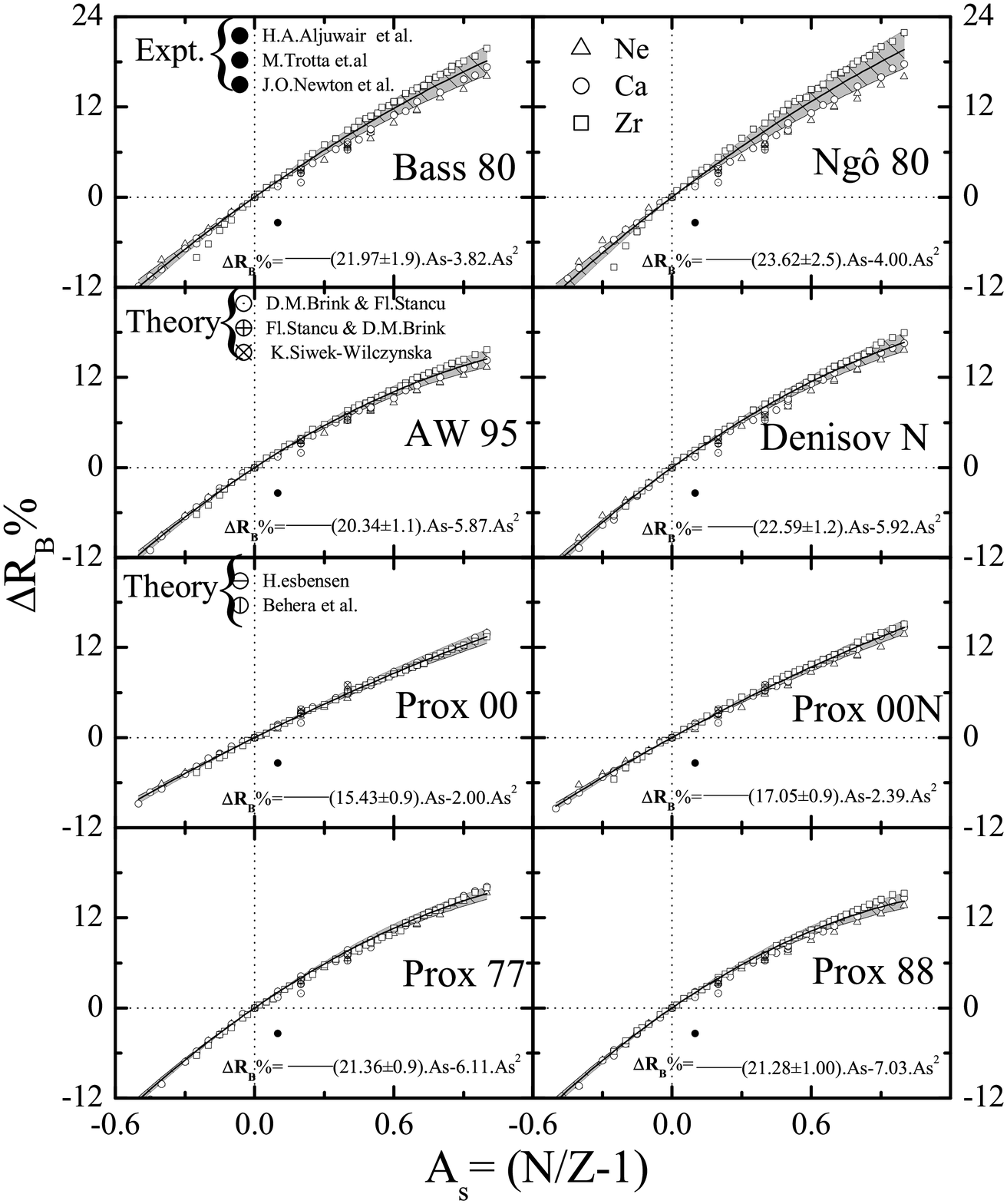}
 \vskip -0cm \caption{The normalized barrier positions $\triangle$ R$_{B}$(\%)
[Define in Eq. (8)] as a function of A$_{s }$(= N/Z-1). We display
the results of our calculations for the collisions of
$^{A1}Ne$+$^{A2}Ne$, $^{A1}Ca$+$^{A2}Ca$ and $^{A1}Zr$+$^{A2}Zr$
series using eight models along with other available theoretical
and experimental values. The theoretical as well as experimental
data reported here is taken from Refs. [28] and [29],
respectively. The shaded areas denotes the deviation from the
cental solid lines.  }\label{fig1}
\end{figure}

\begin{figure}[!t]
\centering
 \vskip 1cm
\includegraphics[angle=0,width=12cm]{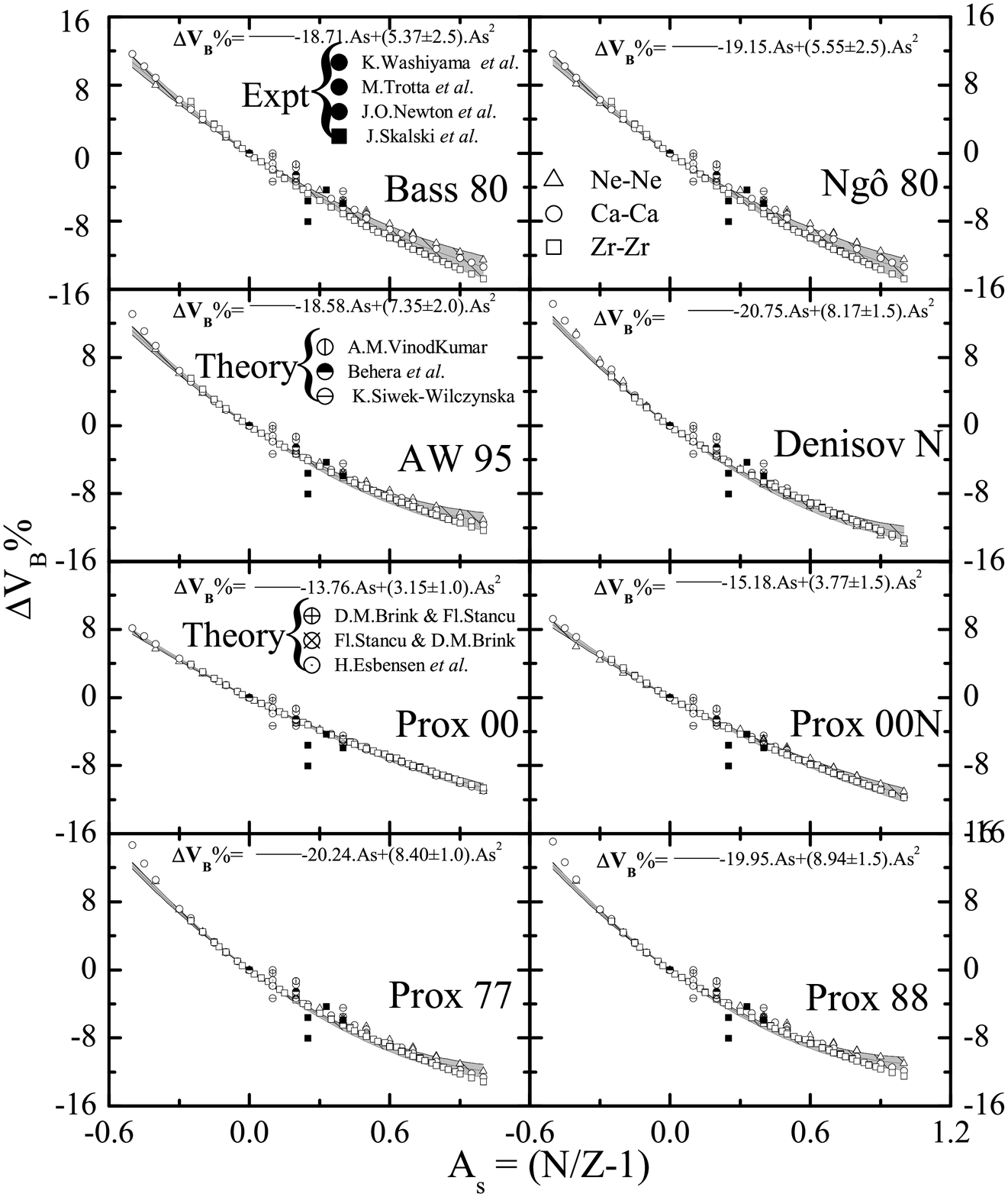}
 \vskip -0cm \caption{Same as Figure 6, but for $\triangle$ V$_{B}$(\%)[Define
in Eq. (9)]}\label{fig1}
\end{figure}

\begin{figure}[!t]
\centering
 \vskip 1cm
\includegraphics[angle=0,width=12cm]{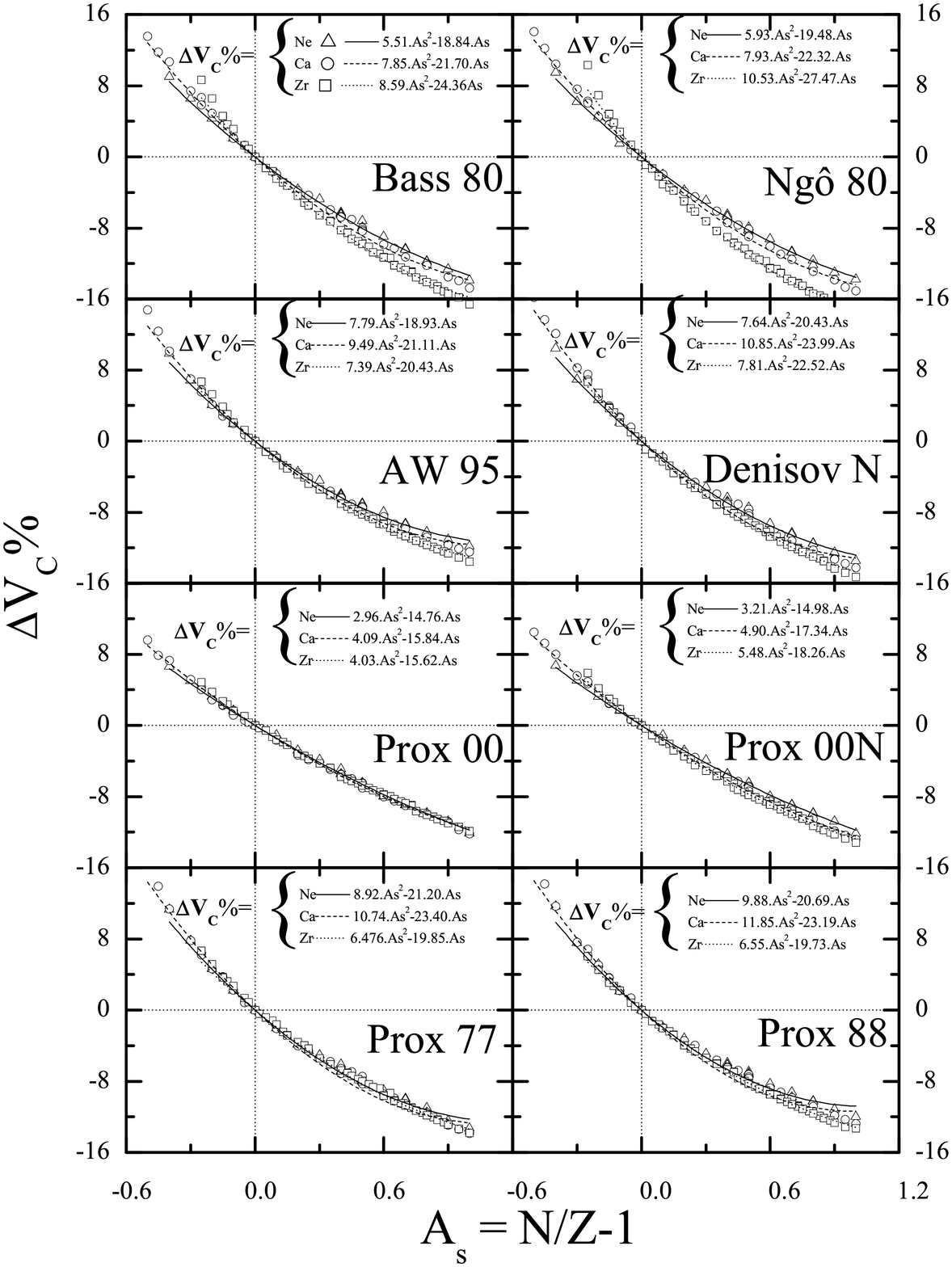}
 \vskip -0cm \caption{The normalized Coulomb barriers potential is plotted as
a function asymmetry A$_{s }$ using eight different potential for
three series. Here non-linear second order fit is
applied.}\label{fig1}
\end{figure}

\begin{figure}[!t]
\centering
 \vskip 1cm
\includegraphics[angle=0,width=12cm]{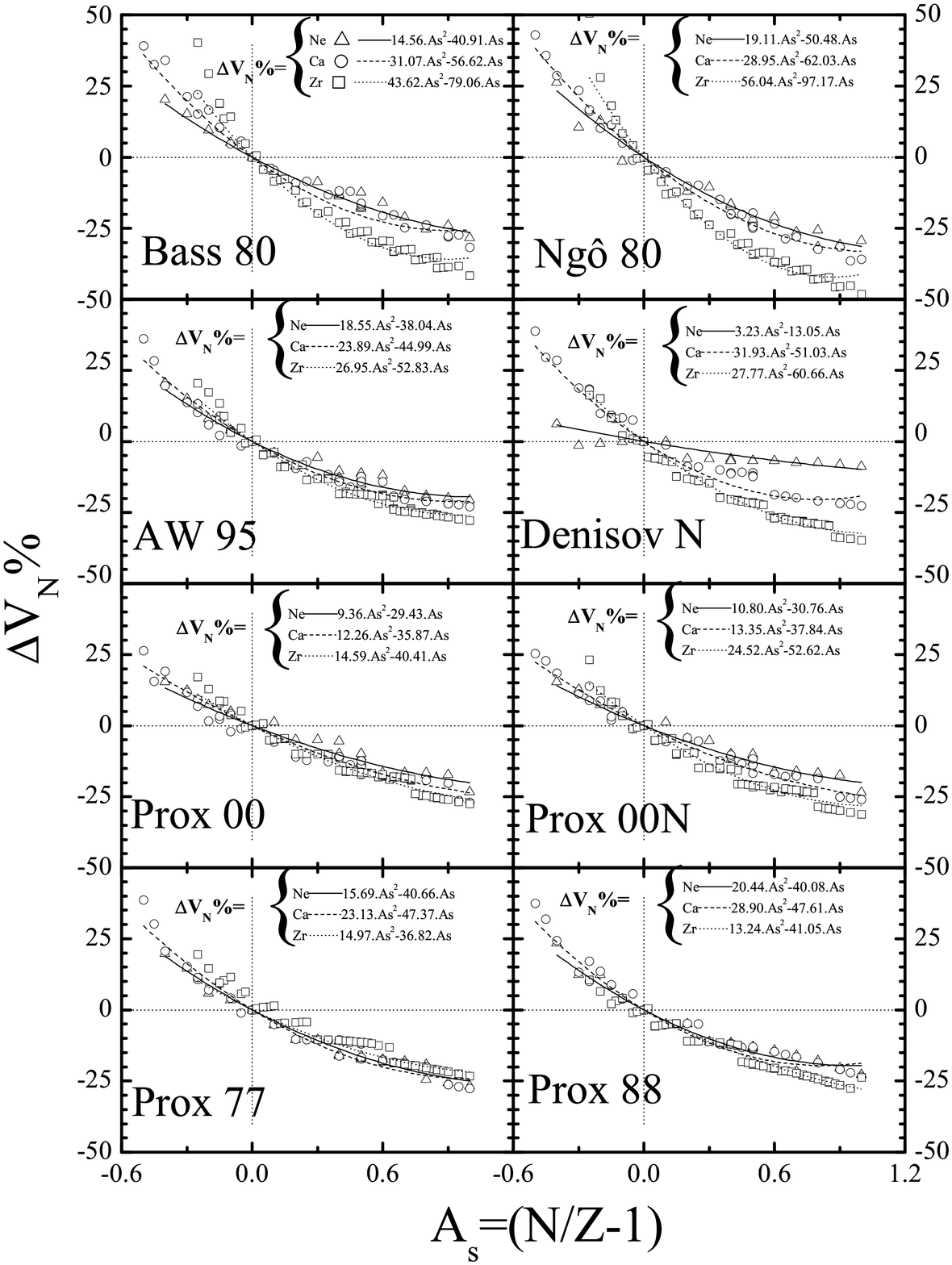}
 \vskip -0cm \caption{ The normalized nuclear potential $\triangle$ V$_{N}$(\%) is plotted as a function of asymmetry
  A$_{s }$ by using eight different potential for three series. Here non-linear second order fit is applied}\label{fig1}
\end{figure}

\par
\section{Summary}
We analyze the fusion of three different series namely, Ne, Ca,
and Zr by covering the wider mass spectrum with N/Z ratio between
0.5 and 2.0. We analyzed the systematic dependence of fusion
barriers on neutron excess and presented a unified second order
non linear quadratic parametrization in fusion barrier heights and
positions with (N/Z-1) using eights isospin dependent
proximity-type models for three different series. Our results are
in good agreement with the available theoretical as well as
experimental results. A linear dependence in the fusion
probabilities is also presented. Further, the enhancement in
fusion cross sections for neutron -rich nuclei due to lowering of
fusion barrier heights is clearly seen, whereas, reverse happen
for proton -rich nuclei. Along with this, our parametrization
pattern is independent of the colliding nuclei as well as model
and isospin content. At near barrier energies, N/Z content plays
dominant role, whereas, the effect is insignificant at above
barrier energies. More experiments are needed to verify our
predications.
\par


\begin{thebibliography}{0}
\bibitem{1} R. K. Puri, {\it et al}, {\it Eur. Phys. J. A} {\bf 23} (2005)
429; \emph{ibid}. A \textbf{8} (2000) 103; \emph{ibid}. A
\textbf{3} (1998) 277; R. K. Puri, {\it et al}, \emph{Phys. Rev.
C} \textbf{43} (1991) 315 ;\emph{ibid}, \emph{Phys. Rev. C}
\textbf{45} (1992) 1837; \emph{ibid},{\bf 315} (1991) C343.
\bibitem{2} R. K. Gupta, {\it et al}, {\it Phys. Rev.
C } {\bf 47} (1993) 561; \emph{ibid} \emph{J. Phys. G:Nucl. Part.
Phys.}{\bf 18} (1992) 1533; \emph{ibid}, \emph{Pramana J.
Phys.}{\bf 32} (1989) 419; R. K. Puri, {\it et al}.,\emph{J. Phys.
G:Nucl. Part. Phys.}{\bf 18} (1992) 903
\bibitem{3} I. Dutt and R. K. Puri,{\it Phys. Rev.
C } {\bf 81} (2010) 044615; \emph{ibid}.{\it Phys. Rev. C } {\bf
81} (2010) 064608; \emph{ibid}.{\it Phys. Rev. C } {\bf 81} (2010)
047601; \emph{ibid}.{\it Phys. Rev. C } {\bf 81} (2010) 064609;
\bibitem{4} R. K. Puri, {\it et al}, {\it Nucl. Phys. A} {\bf 575} (1994)
733; \emph{ibid}.{\it Phys. Rev. C } {\bf 54} (1996) 28; S. Kumar
{\it et al}.,\emph{ibid}.\textbf{58} (1998) 3494; J. Singh {\it et
al}, \emph{ibid} {\bf 62} (2000) 044617;R. K. Puri, {\it et al},J.
Comput. Phys.{\bf 162} (2000) 245; S. Kumar {\it et al}, {\it
Phys. Rev. C } {\bf 78} (2008) 064602.
\bibitem{5} Y. K. Vermani {\it et al},\emph{J. Phys. G:Nucl. Part.
Phys.}{\bf 37} (2010) 015105; \emph{ibid} C \textbf{81} (2010)
014601; \emph{ibid} C \textbf{936} (2009) 0105103;
\emph{ibid}.{\it Phys. Rev. C } {\bf 79} (2009) 064613;
\emph{ibid}.{\it Europhys. Lett. } {\bf 85} (2009) 62001;
\emph{ibid}.{\it Nucl. Phys. A} {\bf 847} (2010) 243;
 A. Sood {\it et al}, {\it Phys. Rev. C } {\bf 79} (2009) 064618;
\emph{ibid}.{\it Phys. Rev. C } {\bf 70} (2004) 034611;
\emph{ibid}.{\it Phys. Rev. C } {\bf 73} (2006) 067602;
\emph{ibid}.{\it Phys. Rev. Lett. B} {\bf 594} (2004) 260;
\emph{ibid}.{\it Phys. Rev. C } {\bf 69} (2004) 054612;
\emph{ibid}. {\it Eur. Phys. J. A} {\bf 30} (2004) 571;
\bibitem{6} S.Gautum {\it et al}, {\it Phys. Rev. C } {\bf 85} (2011) 014603;
 \emph{ibid} \emph{J. Phys. G}{\bf 37} (2010) 085102 ;
 R.chugh. {\it Phys. Rev. C } {\bf 82} (2010) 014603;
 V.Liour {\it et al}.{\it Phys. Rev. Lett. B} {\bf 697} (2011) 512;
S.Goyal {\it et al},{\it Nucl. Phys. A} {\bf 853} (2011) 164; E.
Lehmann {\it et al}, {\it Phys. Rev. C } {\bf 51} (1995) 2113;
{\it progress in Nucl. Particle. Phys} {\bf 30} (1993) 219;
\bibitem{7} M. Trotta. {\it et al},{\it Phys. Rev. Lett } {\bf 84} (2000) 2342;
\bibitem{8} A. M. Vinodkumar. {\it et al},{\it Phys. Rev. C } {\bf 78} (2008) 054608;
W. Loveland. {\it et al},{\it Phys. Rev. C } {\bf 74} (2006)
044607; A. M. Vinodkumar. {\it et al},{\it Phys. Rev. C } {\bf 80}
(2009) 054609;
\bibitem{9}L.F. Canto, P. R. S. Gomes, J. Lubianb, L. C. Chamon,
and E. Crema,{\it Nucl. Phys. A} {\bf 821} (2009) 51;
\bibitem{10}A. M. Stefanini {\it et al}.,{\it Phys. Lett. B } {\bf679} (2009)
95; A. M. Stfanini {\it et al}, {\it Phys. Rev. C } {\bf 73}
(2006) 034606; A. M. Stfanini {\it et al}, {\it Phys. Rev. C }
{\bf 78} (2008) 044607;
\bibitem{11} M. Trotta. {\it et al}, {\it Phys. Rev. C } {\bf 65} (2001) 011601;
\bibitem{12} K. E. Rehm. {\it et al}, {\it Phys. Rev. Lett } {\bf 81} (1998)
3341; E. F. Aguilera.{\it et al}, {\it Phys. Rev. C } {\bf 79}
(2009) 021601; V. N. Kondratyev. {\it et al}, {\it Phys. Rev. C }
{\bf 61} (2001) 044613.
\bibitem{13} A. S. Umar, {\it et al}, {\it Eur. Phys. J. A} {\bf 37} (2008)
245.
\bibitem{14} C. R. Morton. {\it et al},{\it Phys. Rev. Lett } {\bf 72} (1994)
4047; H. Timmers. {\it et al}, {\it Nucl. Phys. A} {\bf 633}
(1998) 421; G. Pollarolo. {\it et al}, {\it Phys. Rev. C } {\bf
62} (2000) 054611; L. Corradi. {\it et al}, {\it Nucl. Phys. A}
{\bf 685} (2001) 37c;
\bibitem{15} M. S. Hussein, {\it et al}, {\it Phys. Rev. C } {\bf 51} (1995)
846; C. H. Dasso, {\it et al}, {\it Phys. Rev. C } {\bf 50} (1994)
R12; N. Takigawa, {\it et al}, {\it Phys. Rev. C } {\bf 47} (1993)
R2470;
\bibitem{16} I. Tanihata. {\it et al}, {\it Phys. Rev. Lett } {\bf 55} (1985)
2676.
\bibitem{17}J. J. Kolata. {\it et al}, {\it Phys. Rev. Lett } {\bf 81} (1998)
4580.
\bibitem{18}X. Y. Sun. {\it et al},{\it Nucl. Phys. A} {\bf 834} (2010) 502;
\bibitem{19} K. Tanaka. {\it et al}, {\it Phys. Rev. Lett } {\bf 104} (2010)
062701; D. Steppenbeck. {\it et al}, {\it Phys. Rev. Lett } {\bf
81} (2010) 014305; M. Takechi. {\it et al}, {\it Nucl. Phys. A}
{\bf 834} (2010) 412; N. Frank. {\it et al}, {\it Nucl. Phys. A}
{\bf 813} (2008) 199; A. V. Dobrovolsky. {\it et al}, {\it Nucl.
Phys. A} {\bf 766} (2006) 1; S.M. Lukyanov. {\it et al}, {\it J.
Phys. G} {\bf 28} (2002) L41; A. Leis- tenschneider. {\it et al},
{\it Phys. Rev. Lett } {\bf 86} (2001) 5442; S. Aoyama, K. Kato,
K. Ikeda, {\it Phys. Rev. C } {\bf 55} (1997) 2379; T. Suzuki.
{\it et al}, {\it Phys. Rev. Lett } {\bf 75} (1995) 3241; L.
Weiss- man, {\it et al}, {\it Phys. Rev. C } {\bf 67} (2003)
054314; T. Ishii, {\it et al}, {\it Eur. Phys. J. A} {\bf 13}
(2002) 15; H. Scheit, {\it et al}, {\it Phys. Rev. C } {\bf 63}
(2001) 014604.
\bibitem{20} K. E. Rehm. {\it et al}, {\it Phys. Rev. Lett } {\bf 81} (1998)
3341.
\bibitem{21} J. R. Brown, {\it et al}, {\it Phys. Rev. C } {\bf 80} (2009)
011306(R); A. Lepine- Szily, {\it et al}, {\it Phys. Rev. C } {\bf
65} (2002) 054318; J. Giovinazzo, {\it et al}, {\it Eur. Phys. J.
A} {\bf 10} (2001) 73; O. N. Malyshev, {\it et al}, {\it Eur.
Phys. J. A} {\bf 8} (2000) 295; B. Blank, {\it et al}, {\it Phys.
Rev. C } {\bf 54} (1996) 572.
\bibitem{22} J. Blocki, J. Randrup, W. J. ¶Swiatecki, and C. F. Tsang, {\it et al}, {\it Ann. Physics(N. Y.) } {\bf 105} (1977)
427.
\bibitem{23}W. D. Myers and W. J. Swiatecki, {\it et al}, {\it Phys. Rev. C } {\bf 62} (2000)
044610;
\bibitem{24}P. Moller and J. R. Nix, {\it et al},{\it Nucl. Phys. A} {\bf 361} (1981) 117;
\bibitem{25}W. Reisdorf,{\it J. Phys. G: Nucl. Part. Phys.} {\bf 20} (1994) 1297;
\bibitem{26}G. Royer and R. Rousseau, {\it et al}, {\it Eur. Phys. J. A} {\bf 42} (2009)
541.
\bibitem{27} V. Y. Denisov. {\it et al}, {\it Phys. Lett. B } {\bf 526} (2002)
315.
\bibitem{28} D. M. Brnk and Fl. Stancu. {\it et al}, {\it Nucl. Phys. A} {\bf 299} (1978) 321; Fl. Stancu and D. M. Brnk, \emph{ibid}{\bf 270} (1976)
236; K. Siwek- Wilczynska and J. Wilczynski, {\it Phys. Rev. C }
{\bf 69} (2004) 024611; H. Esbensen, {\it et al}, {\it Phys. Rev.
C } {\bf 20} (1979) 683; A. M Vinodkumar, {\it et al}, {\it Phys.
Rev. C } {\bf 54} (1996) 791.
\bibitem{29} H. A. Aljuwair, {\it et al}, {\it Phys. Rev. C } {\bf 30} (1984)
1223; M. Trotta. {\it et al}, {\it Phys. Rev. C } {\bf 65} (2001)
011601(R); J. O. Newton, {\it et al}, {\it Phys. Rev. C } {\bf 70}
(2004) 024605; J. Skalski. {\it Phys. Rev. C } {\bf 76} (2007)
044603; K. Washiyama, {\it et al}, {\it Phys. Rev. C } {\bf 78}
(2008) 024610.
\end{thebibliography}
\end{document}